\documentclass[twocolumn]{aastex62}

\usepackage{soul}
\usepackage{amsfonts,amsbsy}
\usepackage{mathrsfs}
\usepackage{bm} 

\usepackage{graphicx} 
\usepackage{dcolumn} 

\usepackage{aas_macros}

\usepackage{xcolor}
\usepackage{color}

\usepackage{url}
\usepackage{multirow}

\usepackage{ulem}
\usepackage{enumerate}
\usepackage{textcmds}
\usepackage{mathtools}
\usepackage{booktabs}
\usepackage{tabularx}
\usepackage{lineno}
\usepackage{enumitem}

\newcolumntype{p}{D{,}{\pm}{-1}}

\renewcommand{\figurename}{Fig.}
\renewcommand{\tablename}{Tab.}

\renewcommand{\arraystretch}{1.5}

\begin{document}

\title[]{Prospects for the detection of the prompt very-high-energy emission from $\rm\gamma$-ray bursts with the High Altitude Detection of Astronomical Radiation experiment}

\correspondingauthor{\\
Yu-Hua Yao, Hao Cai, Tian-Lu Chen, Yi-Qing Guo}

\author{Guang-Guang Xin}
\affiliation{School of Physics and Technology, Wuhan University, Wuhan 430072, P.R. China
}
\affiliation{Key Laboratory of Particle Astrophysics, Institute of High Energy Physics, Chinese Academy of Sciences, Beijing 100049, P.R. China
}

\author{Yu-Hua Yao}
\affiliation{College of Physics, Sichuan University, Chengdu 610064, P.R. China
}
\affiliation{Key Laboratory of Particle Astrophysics, Institute of High Energy Physics, Chinese Academy of Sciences, Beijing 100049, P.R. China
}

\author{Xiang-Li Qian}
\affiliation{
School of Intelligent Engineering, Shandong Management University, Jinan 250357, P.R. China
}

\author{Cheng Liu}
\affiliation{Key Laboratory of Particle Astrophysics,
	Institute of High Energy Physics, Chinese Academy of Sciences, Beijing 100049, P.R. China
}

\author{Qi Gao}
\affiliation{The Key Laboratory of Cosmic Rays (Tibet University), Ministry of Education, Lhasa 850000, Tibet, P.R. China
}

\author{Dan-Zeng-Luo-Bu}
\affiliation{The Key Laboratory of Cosmic Rays (Tibet University), Ministry of Education, Lhasa 850000, Tibet, P.R. China
}

\author{You-Liang Feng}
\affiliation{
	Key Laboratory of Dark Matter and Space Astronomy, Purple Mountain Observatory, Chinese Academy of Sciences, Nanjing 210008, P.R. China
}

\author{Quan-Bu Gou}
\affiliation{Key Laboratory of Particle Astrophysics, Institute of High Energy Physics, Chinese Academy of Sciences, Beijing 100049, P.R. China
}

\author{Hong-Bo Hu}
\affiliation{Key Laboratory of Particle Astrophysics,
	Institute of High Energy Physics, Chinese Academy of Sciences, Beijing 100049, P.R. China
}
\affiliation{University of Chinese Academy of Sciences, 19 A Yuquan Rd, Shijingshan District, Beijing 100049, P.R. China}

\author{ Hai-Jin Li}
\affiliation{The Key Laboratory of Cosmic Rays (Tibet University), Ministry of Education, Lhasa 850000, Tibet, P.R. China
}

\author{Mao-Yuan Liu}
\affiliation{The Key Laboratory of Cosmic Rays (Tibet University), Ministry of Education, Lhasa 850000, Tibet, P.R. China
}

\author{Wei Liu}
\affiliation{Key Laboratory of Particle Astrophysics, Institute of High Energy Physics, Chinese Academy of Sciences, Beijing 100049, P.R. China
}

\author{Bing-Qiang Qiao}
\affiliation{Key Laboratory of Particle Astrophysics,
Institute of High Energy Physics, Chinese Academy of Sciences, Beijing 100049, P.R. China
}

\author{Zhen Wang}
\affiliation{
Tsung-Dao Lee Institute, Shanghai Jiao Tong University, 200240 Shanghai, P.R. China
}

\author{Yi Zhang}
\affiliation{
	Key Laboratory of Dark Matter and Space Astronomy, Purple Mountain Observatory, Chinese Academy of Sciences, Nanjing 210008, P.R. China
}

\author{Hao Cai}
\affiliation{School of Physics and Technology, Wuhan University, Wuhan 430072, P.R. China
}

\author{Tian-Lu Chen}
\affiliation{The Key Laboratory of Cosmic Rays (Tibet University), Ministry of Education, Lhasa 850000, Tibet, P.R. China
}

\author{Yi-Qing Guo}
\affiliation{Key Laboratory of Particle Astrophysics,
	Institute of High Energy Physics, Chinese Academy of Sciences, Beijing 100049, P.R. China
}
\affiliation{University of Chinese Academy of Sciences, 19 A Yuquan Rd, Shijingshan District, Beijing 100049, P.R. China}

\email{yaoyh@ihep.ac.cn, hcai@whu.edu.cn, \\chentl@ihep.ac.cn, guoyq@ihep.ac.cn}

\begin{abstract}

  The observation of very-high-energy (VHE, $\rm >10~GeV$) $\gamma$-ray emission from $\rm \gamma$-ray bursts (GRBs), especially in the prompt phase, will provide critical information for understanding many aspects of their nature including the physical environment, the relativistic bulk motion, the mechanisms of particle acceleration of GRBs and for studying Lorentz invariance violation, etc. For the afterglow phase, the highest energy photons detected to date by the imaging atmospheric Cherenkov telescopes extend to the TeV regime. However, for the prompt phase, years of efforts in searching for the VHE emission has yielded no statistically significant detections. A wide field-of-view (FOV) and large effective area above tens of GeV are essential for detecting the VHE emissions from GRBs in the prompt phase. The High Altitude Detection of Astronomical Radiation (HADAR) experiment has such merits. In this paper, we report the estimates of its expected annual GRB detection rate, which are obtained by combining the performance of the HADAR instrument with the theoretical calculations based on a phenomenological model to generate the pseudo-GRB population. The expected detectable gamma-ray signal from GRBs above the background is then obtained to give the detection rate. In the spectral model, an extra component is assigned to every GRB event in addition to the Band function.  The results indicate that if the energy of the cutoff due to internal absorption is higher than 50 GeV, the detection rate for GRBs for the HADAR experiment is approximately two or three GRBs per year, which varies slightly depending upon the characteristics of the extra component.
\end{abstract}

\section{Introduction}
\label{sec:intro}
  Gamma-ray bursts (GRBs) are among the most luminous events in nature. These events release most of their energy as photons with energies in the range from 30 keV to a few MeV, which is denoted as prompt emission lasting from seconds to thousands of seconds, followed by a long-lasting afterglow with a smaller fraction of the energy radiated in radio, optical, and soft X-rays \citep{M_sz_ros_2006,RevModPhys.76.1143,2015AdAst2015E..22P}. Very-high-energy ($\rm VHE, >10~GeV$) $\rm\gamma$-ray emission from GRBs in both the prompt and the afterglow phases has been proposed in many theoretical works, which is related to various high-energy processes of leptonic and hadronic interactions. Synchrotron and inverse Compton (IC) processes of relativistic electrons and positrons are usually considered as two natural candidates. Detailed calculations of the VHE $\gamma$-ray emissions produced by these processes are presented in \citet{1998ApJ...494L.167P,2000ApJ...537..785D,2007ApJ...671..645A,2009A&A...498..677B}, where, in many cases, the emission spectrum is predicted to extend even up to the TeV range. Ultra-high-energy protons could also be responsible for the prompt emission and may induce distinctive signatures in the GeV-TeV bands by synchrotron emission, muons, or secondary particles injected via photomeson interactions \citep{1998ApJ...509L..81T,1999APh....11..451T,2007ApJ...671..645A,2009ApJ...699..953A}. In the case of hadronic radiation, a much larger amount of energy is required for the engine of GRBs owing to its low radiative efficiency, but \citet{1999APh....11..451T} proposed that a strongly beamed solid opening angle of GRB emission could solve the energy-budget problem. More details regarding the physics of the prompt emission can be found in \citet{2018JApA...39...75I}. The predictions of the VHE $\rm\gamma$-ray emission from GRBs in the afterglow phase can be found in \citet{1998ApJ...501..772P,2001ApJ...548..787S,2001ApJ...559..110Z,2021ApJ...908..225H,2021MNRAS.tmp.1298J}.

\par
  Despite the progress that has been made, many basic questions about GRBs remain open. The observation of the VHE emission during the prompt phase of a GRB would be a great step forward in studying the nature of the central engine \citep{King_2005,2019EPJA...55..132F}, the mechanisms of jet formation \citep{M_sz_ros_1997,2015PhR...561....1K,Ruiz_2016}, the acceleration mechanisms of particles \citep{1995PhRvL..75..386W,1997hecn.confE..16W}, and the environment of the burst’s progenitor \citep{2007NJPh....9...17L,2019EPJA...55..132F}. Known to occur at cosmological distances \citep{1997hecn.confE..16W}, GRBs can also serve as important probes of the extragalactic background light (EBL), \citep{2003AIPC..662..442H}, intergalactic magnetic fields (IGMFs) \citep{Ichiki_2008,2011MNRAS.410.2741T}, and the fundamental nature of space--time \citep{Wei_2016}.

  \par
   A large number of searches for the VHE emission from GRBs in both the prompt and the afterglow phases using a variety of experimental techniques did not lead to detections until recent measurements of the afterglow phase. For the observation of photons of energies in the sub-TeV range, the effective areas of the current space-borne instruments are too small to achieve enough sensitivity. Up to now, the highest-energy photon detected by the Fermi-LAT reaches 95 GeV (or 128 GeV in the rest frame at the redshift of z=0.34), emitted by GRB 130427A in the early afterglow phase \citep{2014Sci...343...42A}. As for the observation of even higher energy $\rm \gamma$ rays, ground-based techniques are available at present (see \citet{2015CRPhy..16..610D, 2016CRPhy..17..663K} for review). Of the two ground-based observation methods, the traditional extensive-air-shower (EAS)-arrays method is relatively insensitive due to its limited ability to distinguish $\rm \gamma$/hadron and its poor resolution, although it has a wide field-of-view (FOV), which makes it more suitable for full-sky searches for GRBs. Despite some indications at low significance, no unambiguous evidence for the detection of the prompt VHE emission from GRBs has been seen in many experiments over years of dedicated efforts \citep{2004ApJ...604L..25A,2007ApJ...655..396H,2007ApJ...666..361A,2017ApJ...843...88A,2017IAUS..324...70B,2019HEAD...1711217H}. The Cherenkov telescope method, of which the majority are the imaging atmospheric Cherenkov telescopes (or IACTs), is inherently more sensitive in both energy and flux. Recently, sub-TeV energy $\rm\gamma$-rays were successfully observed by the MAGIC and the H.E.S.S. telescopes approximately 1 min after the occurance of GRB 190114C and 10 h after the end of the prompt-emission phase of GRB 180720B, respectively \citep{2019Natur.575..459M,2019Natur.575..455M,2019Natur.575..464A}. These events are unique because it was the first time that photons with sub-TeV energies were observed in the afterglow emission phase. Soon after, H.E.S.S. announced the detection of photons with energies up to 4 TeV from GRB 190829A, between 4 and 56 h after the trigger \citep{2021Sci...372.1081H}. MAGIC reported another detection of hundreds of GeV $\rm\gamma$-ray emission from GRB 201216C in the afterglow phase \citep{2020GCN.29073....1M}. Along with the detections of GRB 160821B and GRB 201015A from which over 100 GeV emissions at the $\rm \geq 3 \sigma$ level were announced \citep{2020arXiv201207193M,2020GCN.28659....1B}, there are a total of six GRBs reported with the VHE afterglow emission. The IACTs have, therefore, been making great progress in the observation of the VHE emission in the afterglow phase. It is, however, challenging for IACTs to perform prompt-phase observations because, being pointed observational instruments, they have to slew to the location of the GRB, which means that, apart from in some very unlikely events when the GRB occurs exactly within the region of sky being targeted by the IACT, the telescope will have to slew to the location of the GRB and, therefore, observations of the first seconds of the GRB are impossible.
\par
  It is important to search for the prompt VHE emissions from GRBs using ground-based instruments that can achieve excellent sensitivity and a low energy threshold over a wide FOV. In recent years, the High Altitude Detection of Astronomical Radiation (HADAR) experiment has been proposed. It employs a new observational technique using a large-dimensional refractive water lens as the light collector for observing the Cherenkov light induced by the VHE cosmic-rays (CRs) and $\rm\gamma$-rays in the atmosphere \citep{2017JInst..12P9023C,2019NIMPA.927...46C}. A water-lens prototype with a diameter of R = 0.9 m was constructed and successfully operated in the Yangbajing Observatory (4300 m above sea level, $\rm 90.522^{\circ} E, 30.102^{\circ} N, 606 g/cm^2$) in Tibet, China. The first observation of CRs with a scintillator EAS array in September 2015 was presented in the work of \citet{2017JInst..12P9023C}, and a detailed description of the characteristics and performance of this prototype can be found in \citet{2019NIMPA.927...46C}. A large diameter of 2-m water-lens prototype is planned to be installed in 2021. In this work, the expected annual detection rate of GRBs with the HADAR experiment is presented.

  The rest of this paper is organized as follows. The HADAR observatory and its performance are described in Section 2, followed by a brief discussion of synthetic GRB samples in Section 3. The results and a summary are presented in Section 4.
 \begin{figure*}[!htb]
      \centering
      \includegraphics[width=7cm,height=6cm]{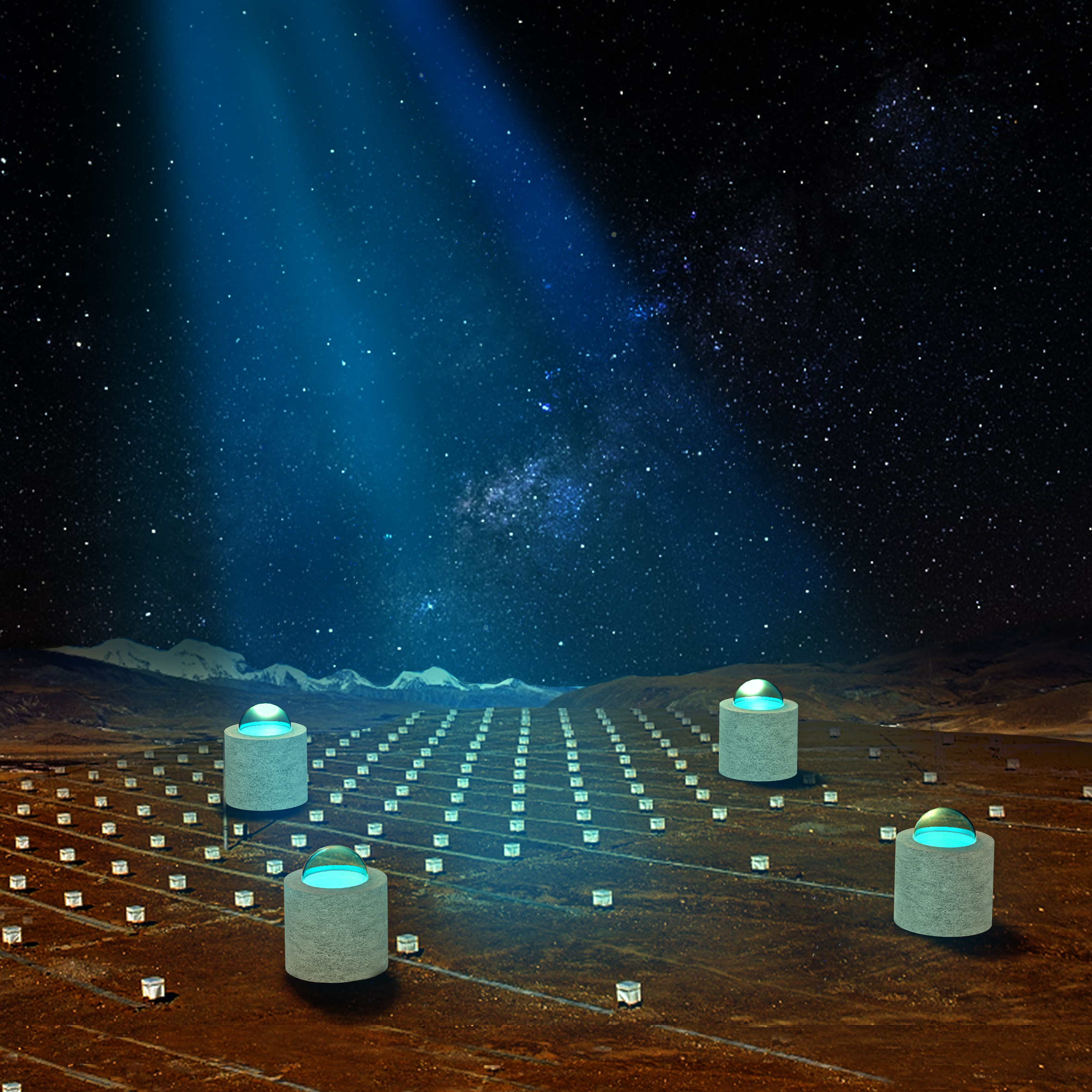}
      \includegraphics[width=7cm,height=6cm]{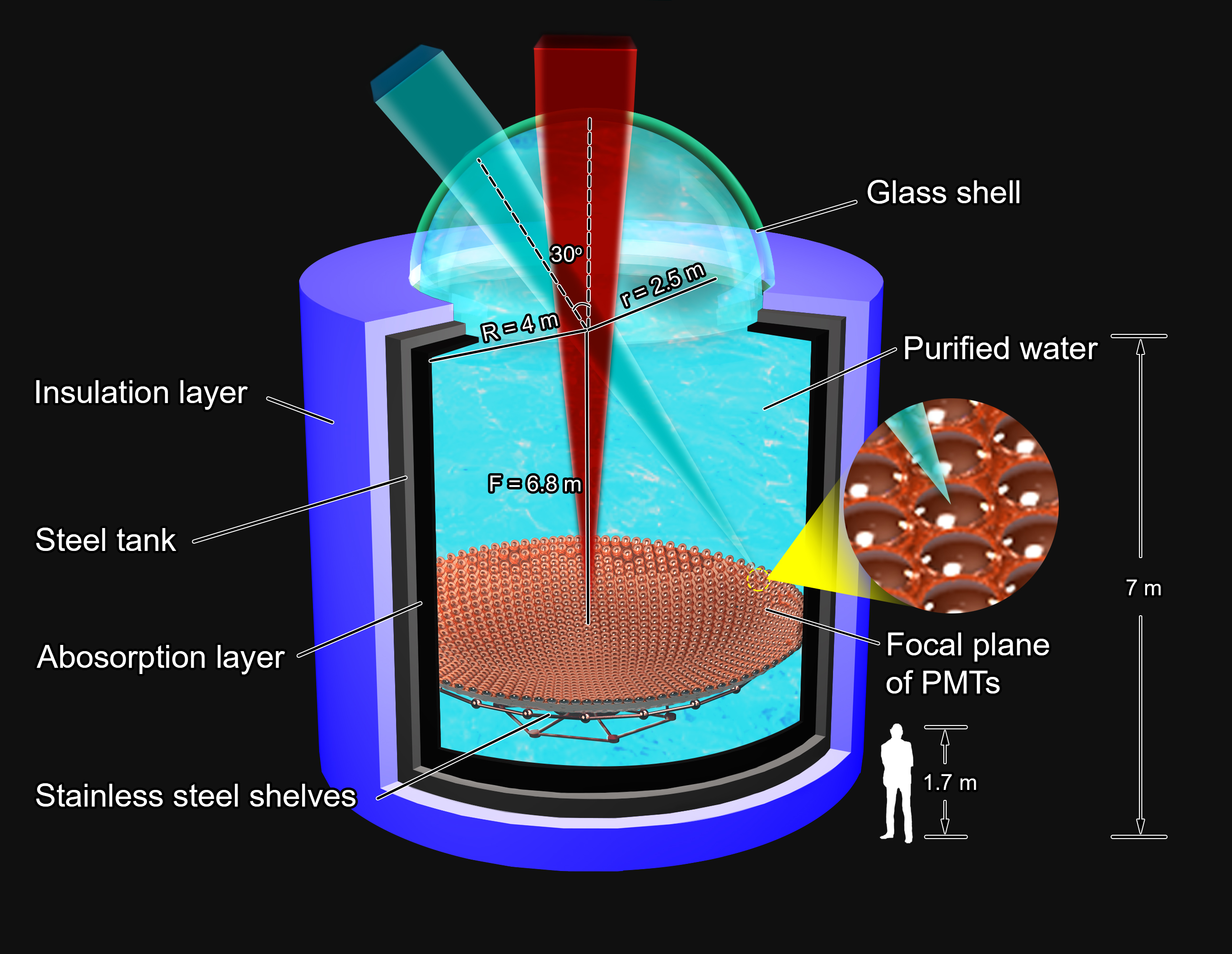}
      \caption{Schematic of the water-lens telescope. Left, layout of the HADAR experiment; right, profile design of each telescope.}
      \label{fig:det}
 \end{figure*}
\section{Experiment}
  \begin{figure*}[!htb]
      \centering
      \includegraphics[height=0.8cm,angle=0,width=0.44\textwidth]{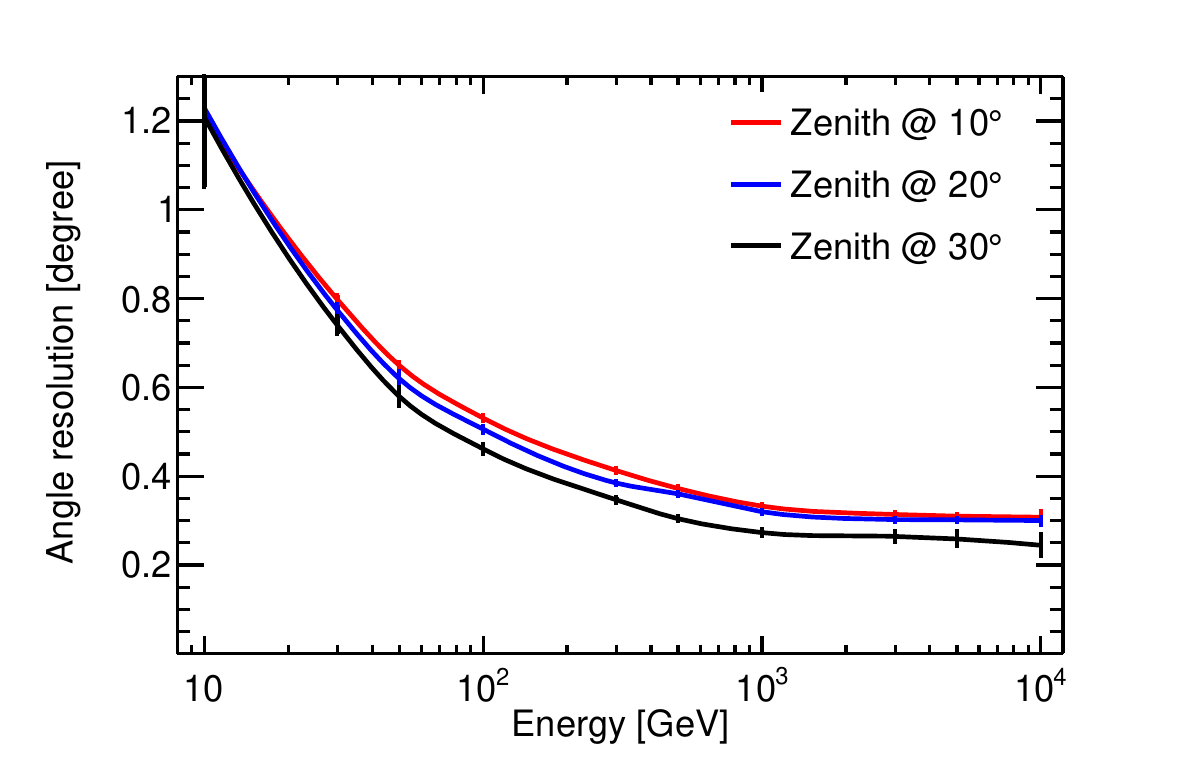}
      \includegraphics[height=0.8cm,angle=0,width=0.44\textwidth]{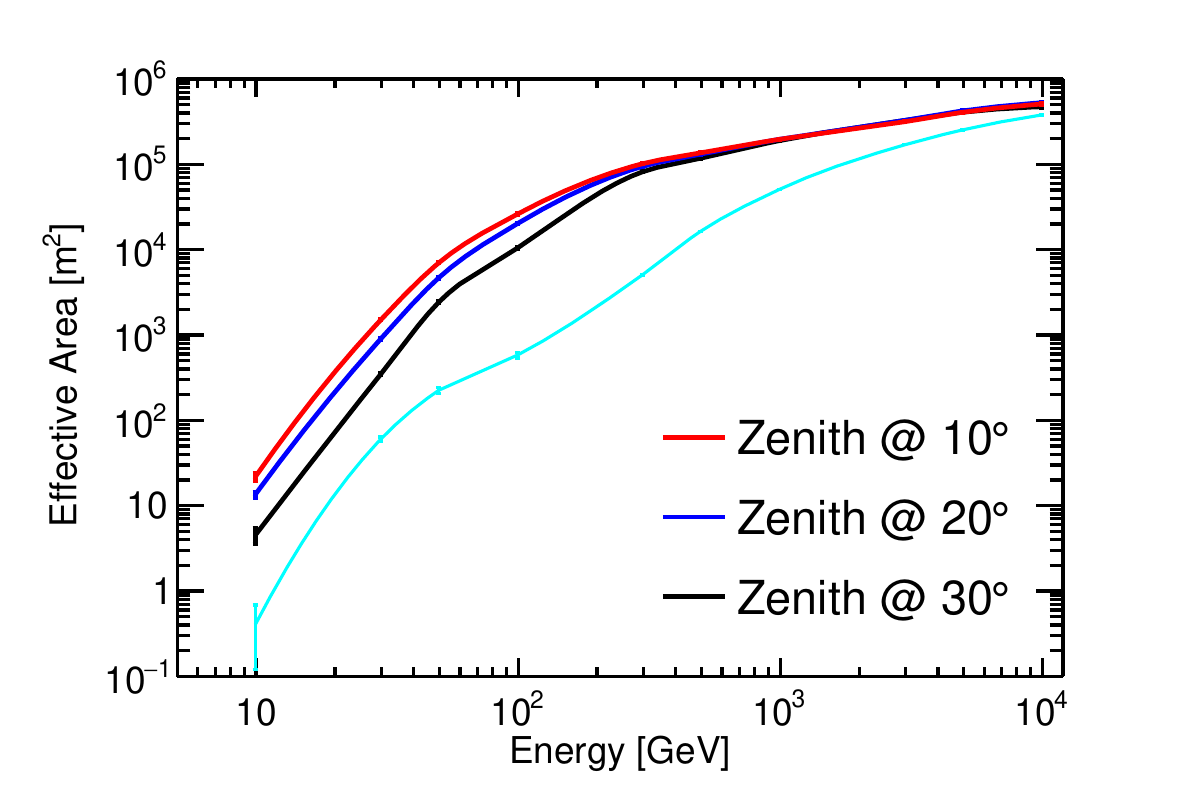}
      \caption{The HADAR performance for $\rm\gamma$-ray detections. Left, angular resolution ($\rm 68\%$ containment radius) at three zenith angles of 10, 20, and 30 deg; right, the effective area. The thin cyan line is the effective area of the cosmic-ray backgrounds.}
      \label{fig:perf}
 \end{figure*}
  The HADAR instrument is configured with four water-lens telescopes to measure the Cherenkov light induced by the $\rm 10~GeV$ to $\rm 100~TeV$ CRs and $\rm\gamma$-rays in the atmosphere. Figure \ref{fig:det} shows the layout of the instrument, in which the four water-lens telescopes are aligned into a square with a side length of 100 m. The smaller cubes are plastic scintillation detectors of the YangBaJing Hybrid Array, which are used to perform joint observations. The right-hand panel of Fig. \ref{fig:det} shows a detailed profile structure of one of the telescopes. It mainly consists of an acrylic spherical cap lens with a diameter of 5 m, a cylindrical tank with a radius of 4 m and a height of 7 m, and a camera with an array of 18,961 PMTs each 5 cm in diameter. The steel tank, containing an absorption layer in the inside wall and thermal insulation material coated in the outer wall, is filled with purified water to reflect radiation emission to the PMTs. The PMTs are placed in the focal plane of the lens and arranged as a series of concentric ring matrixes, supported by a stainless-steel spaceframe.
\par
  Figure \ref{fig:perf} presents the performance of HADAR with different incident rays. As shown on the left, the angular resolution is about a sub-degree in the energy range from approximately 10 GeV to many TeV with almost constant incident angle of the photons. The effective area shown in the right-hand panel increases from tens of square meters to hundreds of thousands of square meters as the energy increases. It is approximately 100,000 square meters at $\rm \sim300~GeV$ for HADAR, which is comparative to that of H.E.S.S. \citep{2015ICRC...34..980H} and MAGIC \citep{2016APh....72...76A}. Note that the observation of GRBs depends not only on the effective area, but also on the FOV. HADAR has a large FOV of approximately 60 deg, which is almost one order of magnitude larger than that of IACTs. The product of the effective area and the solid angle of HADAR is even far larger than that of CTA \citep{2019scta.book.....C}. Owing to these merits, HADAR is suitable for the observation of $\rm \gamma$-ray emission from GRBs, especially in the prompt phase.
\section{Modeling the GRB population}
\label{sec:pop}

  To estimate the capability of HADAR to detect the prompt emission from GRBs, a set of GRB samples was built with Monte Carlo simulations. The method of generating the intrinsic GRB samples is described below. The synthetic GRB populations are simulated based on a set of parameters, space density, intrinsic properties (spectral and temporal), and EBL attenuation.

\subsection{Space density}
  The number of bursts of a specified luminosity at a given redshift is described as \citep{2010MNRAS.406.1944W},
   \begin{equation}\rm
   \label{refer_NLz}
     n(L_p,z)~dlog L_{p}~dz = \phi(L_p)~R(z)~dlog L_{p}~dz,
   \end{equation}
where $\rm\phi(L_p)$ is the isotropic peak luminosity function defined as \citep{2007ApJ...656..306C},

    \begin{eqnarray}\rm~
     \phi(L_p)=a \left\{
     \begin{array}{ll}
\rm~      (L_p/L_p^{\ast})^{\gamma_1}, &\rm~  L_{lower}<L_p<L_p^{\ast},\\
\rm~      (L_p/L_p^{\ast})^{\gamma_2}, &\rm~  L_p^{\ast}<L_p<L_{upper}.
     \end{array}\right.
    \end{eqnarray}
    where a is the normalization constant, the values of other parameters are adopted from \citep{2010MNRAS.406.1944W,2007ApJ...656..306C}, $\rm L_{lower}=10^{50}~erg/s$, $\rm L_{upper}=10^{54}~erg/s$, $\rm L_p^{\ast}=10^{52.5}~ergs/s$, $\rm\gamma_1=-0.17$, and $\rm\gamma_2=-1.44$. $\rm R(z)$ is the differential co-moving rate of the bursts,
   \begin{equation}\rm~
   \label{refer_Rz}
     R(z) = \frac{R_{GRB}(z)}{1+z}~ \frac{dV(z)}{dz},
  \end{equation}
in which one has \citep{2001ApJ...548..522P,2007ApJ...656..306C},
    \begin{equation}\rm
    R_{GRB}(z)=\rho_0~\frac{23~e^{3.4z}}{22+e^{3.4z}}~G(z, \Omega_m, \Omega_\Lambda),
   \label{refer_Rstar}
  \end{equation}
where $\rm G(z, \Omega_m, \Omega_\Lambda)$ is the cosmological term and $\rm\rho_0$ the observed rate of GRBs per differential co-moving volume.

\subsection{Temporal properties}
  The duration of the bursts is approximately estimated by $\rm T_{90}$, which corresponds to the time in which $\rm 90\%$ of the counts arrive. According to \citet{2010MNRAS.403..926G,2012MNRAS.425..514K}, $\rm T_{90}$ is described as,
  \begin{equation}\rm
   T_{90} = (1+z)~ \frac{E_{iso}}{L_{ave}},
  \end{equation}
where isotropic energy $\rm E_{iso}$ is determined by \citet{2012MNRAS.425..514K,2012ApJ...749...99Q},
    \begin{equation}\rm
   \log_{10} (E_{iso}) = 1.1~ \log_{10} (L_p) + 0.56.
  \end{equation}
  The average luminosity $\rm L_{ave}$ is obtained by multiplying the integral luminosity from $\rm 1~keV$ to $\rm 10~MeV$ by a factor of 0.3 according to \citet{2012MNRAS.425..514K}. A comprehensive simulation of the light curve in the prompt phase is complex and is therefore left for future studies.

\subsection{Spectral properties}
  The spectrum is described by the phenomenological Band function \citep{1993ApJ...413..281B},
  \begin{equation}\rm~
    N(E) = J_0 \left \{
    \begin{array}{ll}
  \rm      E^\alpha ~ e^{-E(2+\alpha)/E_p}, &\rm~ E\leq E_{c} \\
  \rm      E^\beta ~ E_{c} ~ e^{\beta-\alpha},&\rm~ E > E_{c},
    \end{array}\right.
  \end{equation}
  where $\rm J_0$ is the normalization constant, $\rm\alpha$ ($\rm\beta$) the low- (high-) energy photon indices, $\rm E_p$ is the peak energy of the spectrum, which is constrained by the $\rm E_p-L_p$ empirical relationship \citep{2009A&A...496..585G,2012ApJ...749...99Q}, and the critical energy $\rm E_c=\frac{\alpha-\beta}{2+\alpha}E_p$. In this work, $\rm\alpha$ and $\rm\beta$ are adopted from the observations of the $\emph{Fermi}$-GBM \citep{2020ApJ...893...46V}.
\par
\begin{figure*}[!htb]
  \centering
  \includegraphics[height=3.3cm,angle=0,width=0.45\textwidth]{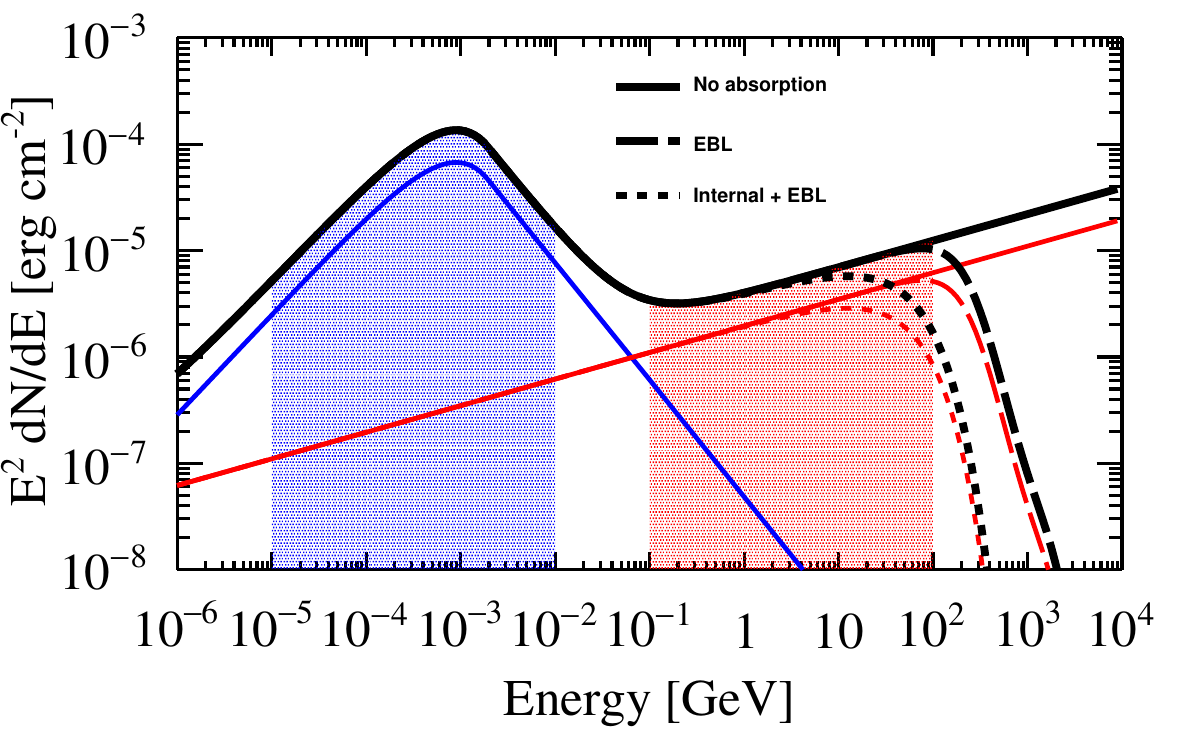}
  \includegraphics[height=5.cm,width=8.cm]{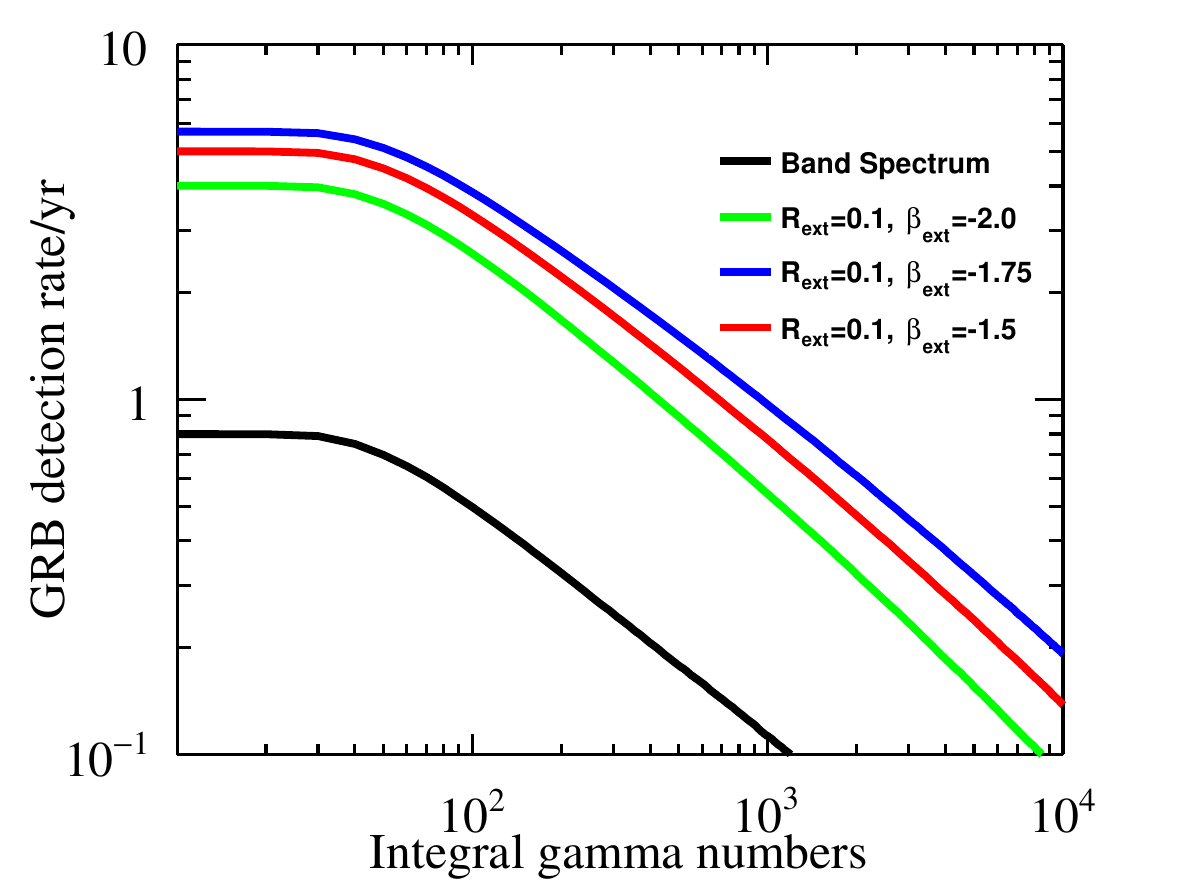}
  \caption{Left, schematically representing the energy spectral distribution of the prompt emission (integrated over $\rm T_{90}$) of GRBs. Blue and red solid lines represent spectra of the Band function and the extra component, respectively. Dashed and dotted lines represent the spectra when taking the EBL absorption \citet{2010ApJ...712..238F} and the internal+EBL absorption into account, respectively; the redshift is adopted as z = 0.42, the cutoff energy is 50 GeV. Black lines are the sum of the Band function and the extra component. Here, values of $\rm\beta_{ext}$ and $\rm R_{ext}$ are -1.75 and 0.1, respectively. $\rm R_{ext}$ represents the ratio of fluences in the energy range of $\rm 100~MeV-100~GeV$ to $\rm 10~keV-10~MeV$, which are the red- and blue-shaded regions, respectively. In order to distinguish the individual components of the emission from the total combined flux, both the Band spectrum and the extra component have been scaled by 0.5. Right, the GRB detection rate of HADAR for different combinations or parameters for the extra component , in which the black line represents the Band spectrum, the green, blue, and red lines represent the combination of the extra + Band components. Here the EBL model of \citet{2010ApJ...712..238F} is considered, and no internal absorption is taken into account.}
  \label{fig:Cartoon}
  \end{figure*}

 In addition to a thermal prompt spectral component which emits just at low energies, an extra component extending to high energies superimposed on the dominant Band function spectral component has been seen in some GRBs. \citep{2009ApJ...706L.138A,2010ApJ...716.1178A,2011ApJ...729..114A}. The exact physical origin of the extra component remains a mystery but the synchrotron self-Compton (SSC) process is likely at play. Additionally, the VHE afterglow emission observed by IACTs also indicates the importance of the SSC process for extremely high-energy emission reported by \citet{2019Natur.575..459M} and \citet{2019Natur.575..448Z}. For simplicity, in this work, it is assumed that the prompt emission spectrum of all GRBs in the synthetic population contains an additional power-law (PL) component, which has spectral index $\rm \beta_{ext}$ and energy ratio $\rm R_{ext}$. Given that the different spectral components may vary among GRBs, $\rm R_{ext}$ is used to obtain the fluence of the extra component by scaling the fluence in the energy regime of the Band function. The left panel of Figure \ref{fig:Cartoon} shows schematically this artificial addition to the Band-function spectrum. $\rm R_{ext}$ represents the ratio of fluences in the energy range from $\rm 100~MeV-100~GeV$ to that from $\rm 10~keV-10~MeV$, which are the red- and blue-shaded regions, respectively. Note that, if the extra component takes the form of a cut-off power law (CPL), or a portion of rather than all GRBs contain the extra component, the results would be between the Band-only and the Band+extra cases.
%

\subsection{$\gamma\gamma$ absorption}
\par

  Low-energy photons within the emission region might strongly attenuate photons of the highest energies through the $\gamma\gamma\rightarrow e^{\pm}$ interaction. The specific spectral breaks due to the pair-production are related with different bulk Lorentz factors and fireball radius   \citep{2018MNRAS.478..749C, 2007ApJ...671..645A}. However, detailed simulations about the evolution process (including the bulk Lorentz factor and the variability time-scale of a single pulse) are outside the scope of this article; besides, the feasible combinations of these parameters (including the bulk Lorentz factor, the fireball radius, and the isotropic equivalent energy) are so many that we cannot exhaust all of them. An exponential cutoff on the energy spectrum is thus introduced to describe the internal absorption when calculating the detection rate. Since for typical burst parameters, the internal shock spectrum cuts off above a threshold energy of $\rm E_{\gamma,th}\sim 10-100 GeV$ and has no internal cutoff with a large enough bulk Lorentz factor \citep{2004ApJ...613.1072R}, the cases of 10 GeV, 30 GeV, 50GeV, 100 GeV, and 1TeV with and without energy cutoff are considered.
\par   
  High-energy photons from cosmological emitters can still suffer the $\gamma\gamma$ interactions with the EBL before reaching the observer. The collective emission of any high-energy emitting cosmological population will exhibit an absorption feature at the highest energies. In this work, several models are adopted in the calculation (see Section 4). Here, the EBL attenuation introduced in the work of \citet{2010ApJ...712..238F} is used by default. The values of the optical depth in a designated energy range at a specific redshift are interpolated.

\section{Results}

  \begin{figure*}[!htb]
  \centering
  \includegraphics[height=5.cm,width=8.cm]{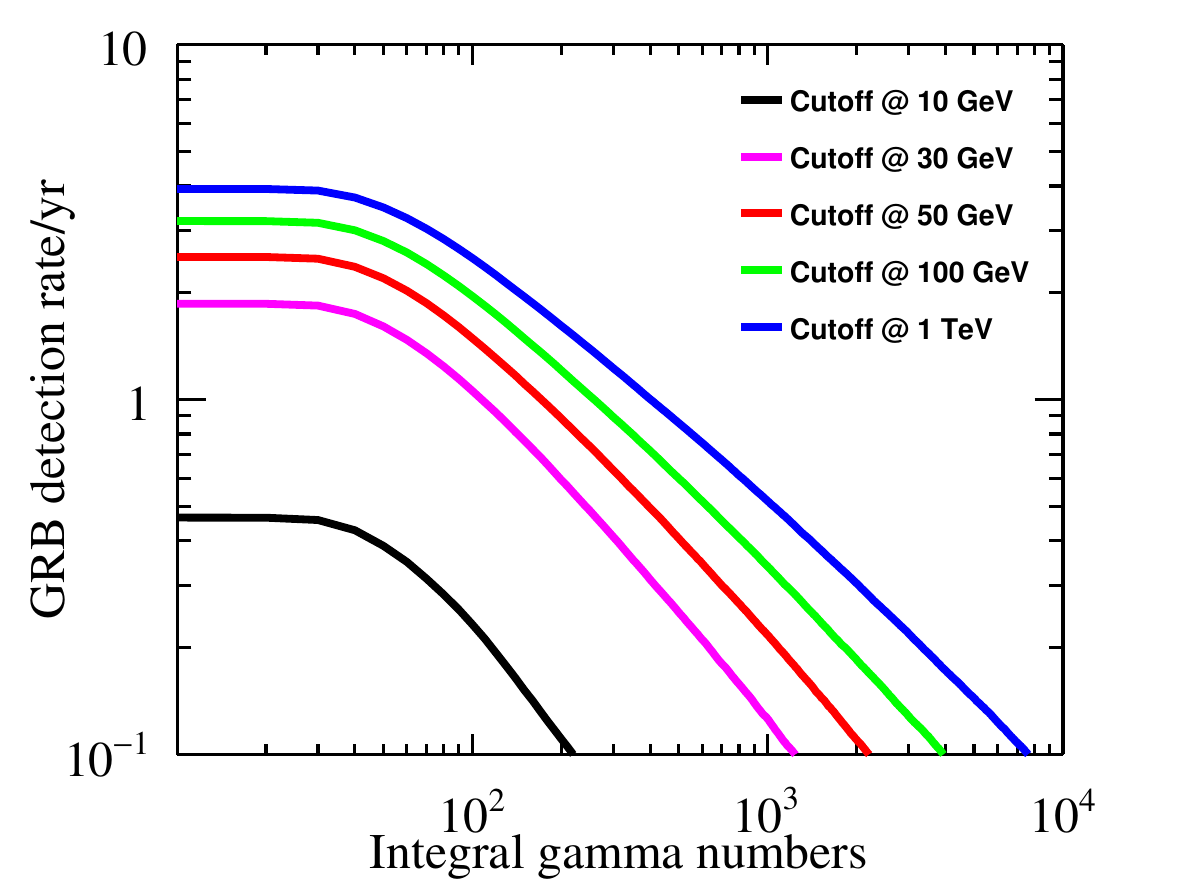}
  \includegraphics[height=5.cm,width=8.cm]{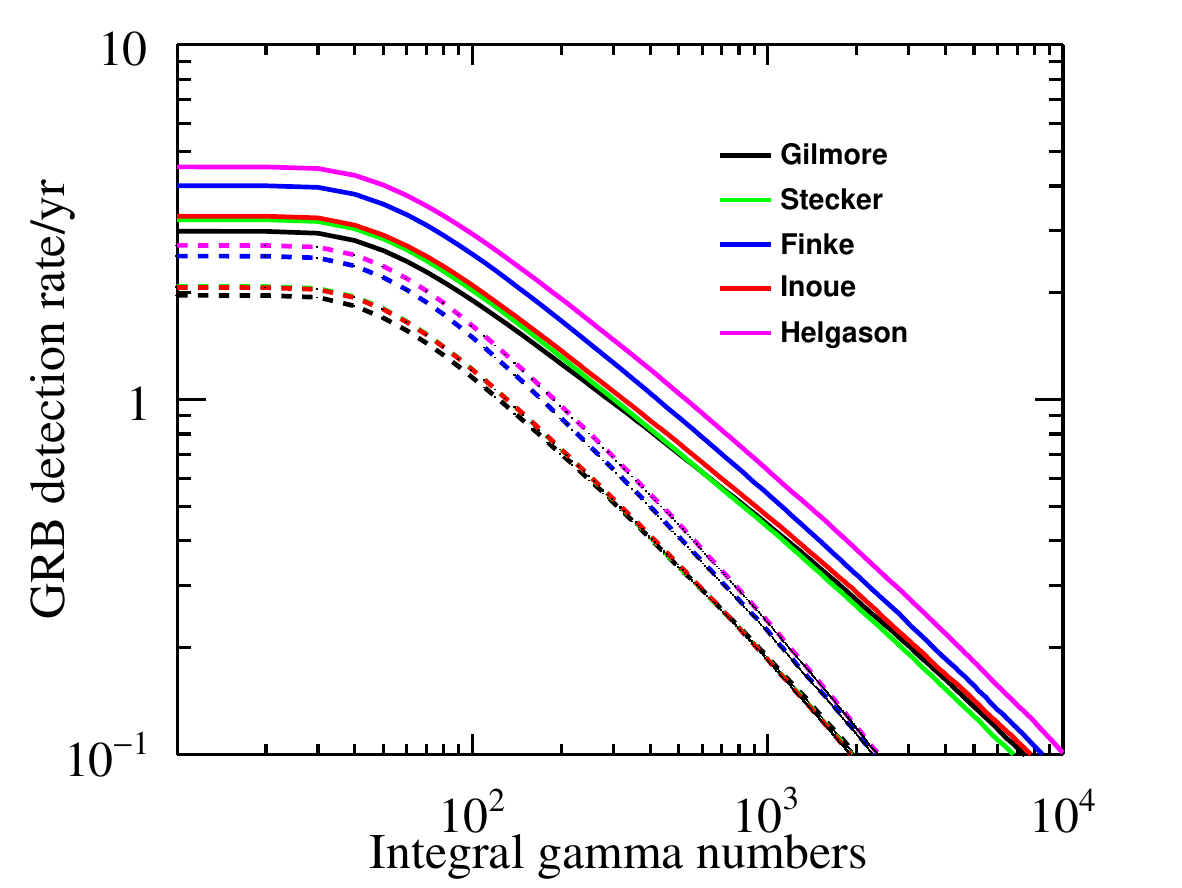}
  \caption{Left, the GRB detection rate with the extra + Band spectra considered for different energy breaks, in the case of $\rm \beta_{ext}=-1.75$, $\rm R_{ext}=0.1$, and the EBL model by \citet{2010ApJ...712..238F}. Right, the GRB detection rate with the extra + Band spectra considered for different EBL models, $\rm \beta_{ext}=-1.75$, $\rm R_{ext}=0.1$. The solid lines are cases without internal absorption, and the dashed lines are the results when the energy spectrum is considered to cut off} at 50 GeV. 
  \label{fig:rateStd}
  \end{figure*}

\par
  When a set of parameters is ready, a pseudo-GRB population is generated and checked with the {\emph{Fermi}}-LAT complete sample of 186 GRBs observed in 10 years. The GRB samples generated match well with the {\emph{Fermi}}-LAT sample. More detailed information about the GRB samples and a comparison can be found in the work of \citet{2020ApJ...900...67K} and \citet{2020ApJ...901..106Y}.
\par
  For the observation of $\rm\gamma$-rays based on ground-based experiments, the overwhelming all-sky cosmic-ray (CR) background is non-negligible. Thus, a $\rm 5~\sigma$ deviation from the background distribution is required to claim the discovery of a GRB. For each pseudo-GRB, the number of signal photons $\rm N_{signal}$ is calculated by,
 \begin{equation}
\rm  N_{signal}=\int_{10 \mathrm{GeV}}^{10 \mathrm{TeV}} S_{\gamma}(E)~ T_{90}~e^{-\tau(E,z)}~A_{e f f}^{\gamma}(E, \theta)~ d E,
\end{equation}
  where $\rm S_{\gamma}(E)$ is the energy spectrum of the pseudo-GRB sample, $\tau(E,z)$ refers to the EBL absorption, and $\rm A_{e f f}^{\gamma}(E, \theta)$ is the effective area of $\rm\gamma$-rays with energy E at zenith angle $\rm\theta$. To estimate the CR background, the same Monte Carlo simulation is run for primary protons based on the Gaisser CR spectrum model \citep{2013FrPhy...8..748G}, obtaining an effective area $\rm A_{e f f}^{p}(E, \theta)$ as a function of energy. The effective area of protons is shown as the thin cyan line in Figure \ref{fig:perf}, and the number of CR backgrounds, $\rm N_{bkg}$, is calculated by,
\begin{equation}
\rm N_{bkg}=\int_{10 \mathrm{GeV}}^{10 T e V} S_{p}(E)~ A_{e f f}^{p}(E, \theta) ~ T_{90} ~ \Omega(E)~d E,
\end{equation}
where $\rm\Omega$ is the angular resolution of HADAR at energy E.
Therefore, the significance of signal photons over the CR background can be estimated with $\rm N_{signal}/\sqrt{N_{bkg}}$.
\par
  As a result, the right-hand panel of Figure \ref{fig:Cartoon} presents the model calculation of the GRB detection rate for HADAR, which is the detected number of GRBs as a function of the number of detected photons. The black, blue, and red lines represent the Band spectra for $\rm R_{ext}$=0.1, and $\rm\beta_{ext}$=-2.0,-1.75,-1.5, respectively. This indicates that the detection rate of HADAR is approximately one per year for the Band spectrum. When the extra component with $\rm R_{ext}=0.1$ is included, the detection rate reaches four, six, and five per year for $\rm \beta = -2.0, -1.75$, and -1.5, respectively. Originating in the cosmos, photons emitted from GRBs might interact with EBL during their journey to Earth, and thus a severe absorption would occur at the highest energies. Therefore, with the same amount of photons, the harder the extra spectrum, the greater the number of high-energy photons and the higher the absorption, leading to less photons remaining with the index -1.5 than with -1.75.
\par

  The results obtained when pair-creation processes within the emission region and during the propagation are considered are presented in Figure \ref{fig:rateStd}. Freezing $\rm R_{ext}$=0.1 and $\rm \beta_{ext} = -1.75$, the detection rate of GRBs with different energy cutoffs are shown in the left-panel of Figure \ref{fig:rateStd}. It can be seen that the internal absorption prevents HADAR from observing the VHE photons to some extent. If the energy cutoff due to the pair creation process within the emission region is higher than 30 GeV, HADAR could detect at least one GRB per year under the EBL model of \citet{2010ApJ...712..238F}. The results in the right-hand panel of Figure \ref{fig:rateStd} show that the different EBL models have a relatively small impact on the detection rate compared with that of the internal absorption. Adopting different EBL absorption models from \citet{2012MNRAS.422.3189G,2012ApJ...761..128S,2010ApJ...712..238F,2013ApJ...768..197I,2012ApJ...758L..13H}, the expected detection rate of HADAR is approximately two or three GRBs per year with or without the 50-GeV energy cutoff.

\section{Conclusions}
  Observations of the VHE emission from GRBs, especially in the prompt phase, are important for our understanding of many related aspects about GRBs. Current instruments are limited due to their poor angular/energy resolution or small FOV/duty cycles. Based on a set of refractive water lens array, HADAR is a uniquely powerful instrument with the advantages of a wide FOV and a large effective area above tens of GeV. In this work, when the internal absorption energy threshold is higher than 30 GeV, we estimate that HADAR has the capacity to observe GRBs with a detection rate of at least one per year. We hope that future experimental observations with HADAR will reveal the physical processes in the prompt phase and allow us to measure the relevant parameters.

\acknowledgments
  This work is supported by the Key R\&D Program of Sichuan Province (Grant $\rm No.~2019ZYZF0001$) and National Natural Science Foundation of China (Grant Nos. 11873005, 12047575, 11705103, 11635011, U1831208, U1632104, 11875264, and U2031110).

\bibliographystyle{aasjournal}
\bibliography{ref}

\end{document}